\documentstyle[prd,aps,epsf]{revtex}

\begin{document}
\bibliographystyle{unsrt}

\iftrue 

\hfill KEK Preprint 97-7

\hfill SMC-PHYS-153

\hfill hep-ph/9704327

\begin{center}
{\large \bf Has the Substructure of Quarks and Leptons Been Found}\\[1mm] 
{\large \bf also by the H1 and ZEUS Detectors at HERA?}\\[4mm]
Keiichi Akama and Kazuo Katsuura\\
{\it Department of Physics, Saitama Medical College,
Kawakado, Moroyama, Saitama 350-04, Japan}\\[2mm]
Hidezumi Terazawa\\
{\it Institute of Particle and Nuclear Studies, HEARO,
Midori-cho, Tanashi, Tokyo 188, Japan}\\[2mm]
\end{center}

\else 
\draft
\title{Has the Substructure of Quarks and Leptons Been Found\\ 
also by the H1 and ZEUS Detectors at HERA?}

\author{Keiichi Akama and Kazuo Katsuura}
\address{Department of Physics, Saitama Medical College,
 Kawakado, Moroyama, Saitama 350-04, Japan}
\author{Hidezumi Terazawa}
\address{\it Institute of Particle and Nuclear Studies, HEARO,
 Midori-cho, Tanashi, Tokyo 188, Japan}
\maketitle

\fi 

\begin{abstract}
The significant excess of events 
	recently found by the H1 and ZEUS Collaborations at HERA 
	in the deep-inelastic $e^+p$ scattering 
	for high momentum-transfers squared $Q^2>15000 {\rm GeV}^2$ 
	over the expectation of the standard model can be explained 
	by either one of the following possible consequences 
	of the substructure of quarks and leptons:
1)  production of leptoquarks ($\Phi $),
2)  exchange of an excited $Z$ boson ($Z'$), 
3)  intrinsic form factors of quarks (or leptons), 
4)  intrinsic anomalous magnetic moments of quarks, 
5)  production of excited quarks ($q^*$),
and
6)  that of excited positrons ($e^{*+}$).
The masses of these new particles are estimated to be 
	$m_{\Phi } \cong 280-440 {\rm GeV}$, $m_{Z'} \cong 200-250 {\rm GeV}$, 
	$m_{q^*} \cong 120-140 {\rm GeV}$, 
	and $m_{e^*} \cong 300-370 {\rm GeV}$, 
	although the possibilities 2) and 5) are excluded 
	by the currently available experimental constraints. 
\vskip 12pt\noindent PACS number(s): 12.60.Rc, 13.60.Hb, 14.80.-j
\vskip 12pt
\end{abstract}

Very recently, the H1 \cite{1} and ZEUS \cite{2} Collaborations at HERA have 
	reported their data on the deep-inelastic $e^+p$ scattering 
	with a significant excess of events over the expectation 
	of the standard model of electroweak and strong interactions 
	for high momentum-transfers squared $Q^2 > 15000 {\rm GeV}^2$, 
	which may indicate a sign for new physics beyond the standard model.
There have already appeared many phenomenological analyses \cite{3} 
1	in which various interpretations of the excess are presented.  
They are either by productions of leptoquarks, ``leptogluons'', 
	and squarks, or by contact interactions of quarks and leptons.  
In our previous Rapid Communication \cite{4}, 
	we have tried to explain the significant excess 
	earlier found by the CDF Collaboration \cite{5} 
	in the inclusive jet cross section for jet transverse energies $E_T
 	 \geq  200 {\rm GeV}$ over current QCD predictions 
	either by possible production of excited bosons 
	(excited gluons, weak bosons, Higgs scalars, etc.) 
	or excited quarks and have concluded that excited quarks 
	whose masses are around $500 {\rm GeV}$ are excluded 
	while excited bosons whose masses are around 2 TeV 
	are allowed for the explanation.  
The purpose of this Rapid Communication is to report our similar trial for 
	explaining the new excess found by the H1 and ZEUS Collaborations 
	by either one of the following possible consequences 
	of the substructure of quarks and leptons \cite{6,7,8}:
1)  production of leptoquarks ($\Phi $),
2)  exchange of an excited $Z$ boson ($Z'$),
3)  intrinsic form factors of quarks (or leptons),
4)  intrinsic anomalous magnetic moments of quarks,
5)  production of excited quarks ($q^*$),
and
6)  that of excited positrons ($e ^{*+}$).

The minimal composite model of quarks and leptons \cite{6,7,8} 
	consists of an isodoublet of subquarks, $w_i (i=1,2)$ 
	with the charge of $\pm 1/2$, and a color quartet of subquarks, 
	$C_{\alpha } (\alpha = 0,1,2,3)$ 
	with the charge of $-1/2$ (for $\alpha = 0$) 
	and $+1/6$ (for $\alpha = 1,2,3$).  
The quarks ($q$) and leptons ($l$) are taken as composites of subquarks 
	consisting of $w_i$ and $C_\alpha $ 
	while the weak bosons ($W^\pm $ and $Z$), 
	the gluons ($G^a$, $a=1,2,...,8$), 
	the Higgs scalars ($\phi ^+, 
	\phi ^0$) [and even the photon ($\gamma $)] 
	can also be taken as composites of a subquark and an antisubquark 
	such as $w_i$ and $\overline {w_j}$ 
	or $C_\alpha $ and $\overline {C_\beta }$.  
In this model, we expect that there may appear 
	not only exotic states and excited states of the fundamental bosons 
	such as the leptoquarks ($\Phi ^{\pm 2/3}$) consisting of 
	$\overline {C_0}(C_0)$ and 
	$C_\alpha(\overline {C_\alpha}) (\alpha = 1,2,3)$ 
	and the excited $Z$ boson ($Z'$) consisting of 
	$w_i (C_\alpha )$ and $\overline {w_i} (\overline {C_\alpha })$, 
	but also excited states of the fundamental fermions such as 
	the excited quarks ($q^*$) consisting of 
	$w_i$ and $C_\alpha  (\alpha = 1,2,3)$ 
	and the excited leptons ($l^*$) consisting of 
	$w_i$ and $C_0$ \cite{9,10}.  
Also, we expect that quarks (and leptons) 
	may have intrinsic form factors \cite{11} 
	and intrinsic anomalous magnetic moments \cite{12} 
	due to the substructure.  
Note that the existence of leptoquark gauge bosons has been predicted 
	in the grand unified gauge theories of 
	strong and electroweak interactions \cite{13} 
	but with extremely large masses 
	of the grand unification energy scale ($> 10^{15} {\rm GeV}$).  
Also note that the existence of additional extra $Z$ bosons 
	($Z_R$, etc.) has been predicted 
	in many extensions of the standard model \cite{14} 
	with fairly large masses ($> 500 {\rm GeV}$).  
Here, however, we concentrate on the possibility 
	of relatively low energy scale of the order of 1 TeV 
	for compositeness, which has recently been anticipated 
	rather theoretically \cite{15} or 
	by precise comparison between currently available experimental data 
	and calculations in the composite models 
	of quarks and leptons \cite{16,17}.  
In what follows, we shall check the above mentioned possibilities 1)--6) 
	one by one.  

Let us first define some kinematical variables 
	for the deep inelastic scattering of 
	$e^+ + p \rightarrow e^++$ anything.  
Let $P$, $k$, and $k'$ be the momenta of the initial proton, 
	the initial positron, and the final positron, respectively.  
Then, the Mandelstam variables are defined as 
	$s = (P+k)^2$ and $t = (k-k')^2(\equiv-Q^2)$, 
	and the scaling variables as $x = |t| /2P\cdot q$ 
	and $y = P \cdot q / P \cdot k$ for $q = k-k'$, 
	which have an approximate relation of $sxy \cong |t|$ 
	for $s \gg m_p^2$ where $m_p$ is the proton mass.  
In the parton model, the differential cross section 
	for $e^+ p$ scattering is given by 
\begin{eqnarray}
	d \sigma = \sum _a f_a(x)dx d \widehat \sigma _a,
\label{eq:1}
\end{eqnarray}
where $f_a(x)$ is the distribution function for the parton $a$ 
	and $d \widehat \sigma _a$ is the differential cross section 
	for the subprocess of $e^+ a \rightarrow e^+ a$.  
As for the parton distribution functions (PDFs), 
	we use those of Gl\"{u}ck-Reya-Vogt \cite{18}.  
We expect that the ratio of our composite model calculation 
	to the standard model one may not so much depend on PDFs.  
For definiteness, the differential cross section 
	for the subprocess $e^+ a \rightarrow e^+a$ is given by 
\begin{eqnarray}
	d\widehat \sigma _a / dt = 16 \pi \left[ 
		(1-y)^2 (|P_{LL}|^2 + |P_{RR}|^2) 
	        +       (|P_{LR}|^2 + |P_{RL}|^2)\right], 
\label{eq:2}
\end{eqnarray}
with
\begin{eqnarray}
	P_{XY}
 	= {-e^2 Q_a   \over t} 
	+ {g_X^e g_Y^a\over t-m_Z^2} \ \  (X,Y = L,R),
\label{eq:3}
\end{eqnarray}
where $e$ is the electromagnetic coupling constant ($e^2=4\pi \alpha $), 
	$Q_a$ is the charge of $a$, 
	$g_X^\psi $ is the coupling constant of $Z$ to the fermion $\psi $ 
	($\psi = e$ or $q$) with the chirality $X$ ($X=L$ or $R$), 
	and $m_Z$ is the $Z$ boson mass.

The first possibility is the production of leptoquarks.  
Since the leptoquark $\Phi ^{+2/3}$ consisting of 
	$\overline {C_0}$ and $C_{\alpha } (\alpha =1,2,3)$ 
	can be produced by annihilation of $e^+$ 
	(consisting of $\overline {w_2}$ and $\overline {C_0}$) 
	and $d$ [consisting of $w_2$ and $C_\alpha (\alpha =1,2,3)$] 
	but neither by annihilation of $e^+$ and $u$ or 
	by collision of $e^-$ (consisting of $w_2$ and $C_0$) 
	and any quarks [consisting of $w_i (i=1,2)$ 
	and $C_{\alpha }$], we expect the possible excess 
	due to the production of $\Phi $ in $e^+p$ scatterings 
	but not in $e^-p$ scatterings.  
Let us assume the scalar leptoquark $\Phi _i (i=1,2,3)$ for simplicity.  
Then, the effective interaction Lagrangian is given by
\begin{eqnarray}
	L_\Phi  = f_\Phi   \overline l \Phi _i^\dagger  d_i + {\rm H.c.},
\label{eq:4}
\end{eqnarray}
where $f_\Phi $ is the coupling constant.  
The differential cross section for inelastic $e^+p$ scatterings 
	including the effect of $\Phi $ production can be calculated 
	by using Eqs.~(\ref{eq:1}) and (\ref{eq:2}) 
	with the following terms added to $P_{LR}$ and $P_{RL}$:
\begin{eqnarray}
	\Delta P_{LR} = \Delta P_{RL} 
	= {		f_\Phi ^2
\over 
		4(\widehat s - m_\Phi ^2 + im_\Phi \Gamma _\Phi )    },
\label{eq:5}
\end{eqnarray}
where $\widehat s (=xs)$ is the invariant mass squared of $e^+d$ channel, 
	$m_{\Phi }$ is the $\Phi $ mass, and $\Gamma _{\Phi }$ is the decay width.  
If $\Phi $ decays predominantly into two-body channels of $\Phi \rightarrow l \overline {q}$, 
	the decay width is approximated as 
\begin{eqnarray}
	\Gamma _\Phi  \cong 
	{f_\Phi^2 m_\Phi \over 8 \pi }
	\left[ 5 + \theta (m_\Phi ^2 - m_t^2) 
              \left( 1- {m_t^2\over m_\Phi ^2}\right) ^{1/2}\right],
\label{eq:6}
\end{eqnarray}
where $m_t$ is the top-quark mass.  
The H1 and ZEUS experiments \cite{1,2} altogether 
	have observed 10 events for $|t| \geq  20000 \rm GeV^2$ 
	while they have expected 4.08 events in the standard model.  
We estimate the expected number of events in the model to be 
	$n_{th}=n_{SM}\sigma/\sigma_0$
	where $n_{SM}$ is the expected number of events 
	in the standard model given in Refs.\cite{1,2},
	and $\sigma$ and $\sigma_0$ are the integrated cross sections
	with and without the composite-model effects, respectively.
The upper (lower) bound of $m_\Phi$ at the confidence level $p$
	is given by the value for which the Poisson probability 
	to find as many as or more (less) than the actually ovserved 
	number of events becomes $(1-p)/2.$ 
Therefore, 
	we can obtain the bounds on the $\Phi $ mass 
	as $291 {\rm GeV} \leq  m_\Phi  \leq  444 {\rm GeV}$  at 95\% C.L.
	for $f_\Phi =g$  
	(the gauge coupling constant of $SU(2)_L$ in the standard model). 
On the other hand, 
	by knowing that 313 events are observed
	in the whole region of $|t| \geq  5000 {\rm GeV}^2$,
	where 313 events are expected in the standard model,
	we can obtain the lower bound on $m_\Phi $ 
	as $m_\Phi  \geq  271 $GeV at 95\% C.L. for $f_\Phi =g$.
In Fig.\ \ref{f1}, the ratio ($r$) of the differential cross section 
	for inelastic $e^+p$ scattering including the effect 
	of $\Phi $ production for $m_{\Phi } = 271$GeV and $f_\Phi =g$ 
	to that in the standard model 
	is compared with the corresponding H1 experimental result \cite{1}.

The second possibility is the exchange of an extra $Z$ boson ($Z'$).
For simplicity, let us assume that $Z'$ 
	behaves as $\overline {w_L} \gamma _{\mu } w_L$ 
	or a singlet of $SU(2)_L$.  
Then, the effective interaction Lagrangian is given by
\begin{eqnarray}
	L_{Z'} = 
   	f_{Z'} \sum _{\psi } 
	\overline {\psi _L} \gamma ^{\mu } Z'_{\mu } \psi _L,
\label{eq:7}
\end{eqnarray}
where $f_{Z'}$ is the coupling constant and the summation is 
 	over all quarks and leptons, {\it i}.{\it e}.
	~$\psi = q_i (i=1,2,3)$ and $l$.  
The differential cross section for inelastic $e^+p$ scattering 
	including the effect of $Z'$ exchange 
	can be calculated by using Eqs.~(\ref{eq:1}) and (\ref{eq:2}) 
	with the following term added to $P_{LL}$:
\begin{eqnarray}
	\Delta P_{LL} = {
		  f_{Z'}^2
\over 
		  t-m_{Z'}^2    },
\label{eq:8}
\end{eqnarray}
where $m_{Z'}$ is the $Z'$ mass.  
For having 10 events for $|t| \geq  20000 {\rm GeV}^2$, 
	the H1 and ZEUS results \cite{1,2} 
	indicate $m_{Z'} \leq  248 $GeV 
	for $f_{Z'}=g$ (95\% C.L.) 
	while for having 313 events for $|t| \geq  5000 {\rm GeV}^2$, 
	they suggest $m_{Z'} \geq  199 $GeV 
	for $f_{Z'}=g$ (95\% C.L.).
In Fig.\ \ref{f1}, the ratio $r$ of the differential cross section 
	for inelastic $e^+p$ scattering including the effect 
	of $Z'$ exchange for $f_{Z'}=g$ 
	and $m_{Z'} = 199 $GeV to that in the standard model 
	is compared with the H1 result \cite{1}.  

The third possibility is the effect of intrinsic form factors 
	of quarks (or leptons).  
For simplicity, let us assume that 
	both the $\gamma _{\mu }$ vertex of the photon and quarks (or leptons) 
 	and that of the $Z$ boson and quarks (or leptons) be modified 
	into $\gamma _{\mu } F(t)$ and that the form factor $F(t)$ 
	is approximately parametrized as
\begin{eqnarray}
	F(t) \cong 1+ {|t|/\Lambda _F^2} 
\ \ \   {\rm for}\ \ |t| \ll \Lambda _F^2,
\label{eq:9}
\end{eqnarray}
	where $\Lambda _F$ is the mass scale parameter.  
Then, the differential cross section for inelastic $e^+p$ scattering 
	including the effect of quark (or lepton) intrinsic form factors 
	can be calculated by using Eqs.~(\ref{eq:1}) and (\ref{eq:2}) 
	simply multiplied by the factor of $|F(t)|^2$.  
For having 10 events for $|t| \geq  20000 {\rm GeV}^2$, 
	the H1 and ZEUS results \cite{1,2} indicate 
	$\Lambda _F \leq  546 $GeV (95\% C.L.) 
	while for having 313 events for $|t| \geq  5000$GeV$^2$, 
	they suggest $\Lambda _F \geq  365 $GeV (95\% C.L.).
In Fig.\ \ref{f1}, the ratio $r$ including the form factor effect 
	for $\Lambda _F = 365 $GeV is compared with the H1 result \cite{1}.  

The fourth possibility is the effect 
	of intrinsic anomalous magnetic moments of quarks.
Since the effect of intrinsic anomalous magnetic moments of leptons 
	due to the possible substructure of leptons, if any, 
	is not only extremely small but also hard to be discriminated 
	against radiative corrections to the cross sections, 
	we simply ignore it.  
For simplicity, we assume the universality 
	for the intrinsic anomalous magnetic moments of quarks, 
	which is parametrized by the effective interaction Lagrangian of
\begin{eqnarray}
	L_A = 
	\sum _q {eQ_q\over 4 \Lambda _A} 
	\overline {q_i} \sigma _{\mu \nu } F^{\mu \nu }q_i,
\label{eq:10}
\end{eqnarray}
	where $Q_q$ is the quark charge and $\Lambda _A$ 
	is the mass scale parameter.  
The differential cross section for inelastic $e^+p$ scattering 
	including the effect of quark intrinsic anomalous magnetic moments 
	can be calculated by using Eqs.~(\ref{eq:1}) and (\ref{eq:2}) 
	with the following term added to $d \widehat \sigma _a/dt$:
\begin{eqnarray}
	\Delta \left( {d\widehat \sigma _a\over dt}\right)  
	= {\pi \alpha Q_q^2 (1-y) \over  \Lambda _A^2|t|}.
\label{eq:11}
\end{eqnarray}
For having 10 events for $|t| \geq  20000 {\rm GeV}^2$, 
	the H1 and ZEUS results \cite{1,2} 
	indicate $\Lambda _A \leq  171 {\rm GeV}$ (95\% C.L.) 
	while for having 313 events for $|t| > 5000 {\rm GeV}$, 
	they suggest $\Lambda _A \geq  119 {\rm GeV}$ (95\% C.L.).
In Fig.\ \ref{f1}, the ratio $r$ including the effect 
	for $\Lambda _A=119 {\rm GeV}$ 
	is compared with the H1 result \cite{1}.  

The fifth possibility is the production of excited quarks 
	through the subprocess of $e^+q \rightarrow e^+q^*$.  
For simplicity, we assume the universality for all quarks.  
The effective interaction Lagrangian is then given by 
\begin{eqnarray}
	L_{q^*} 
	= \sum _q {eQ_q\lambda _{q^*}\over 2m_{q^*}} 
	    \overline {q_i^*} \sigma ^{\mu \nu } F_{\mu \nu } q_i 
             + {\rm H.c.},
\label{eq:12}
\end{eqnarray}
	where $m_{q^*}$ is the excited quark mass and 
	$\lambda _{q^*}$ is a constant parameter.  
Let us define the parton momentum fraction 
	$\xi $ by $\xi = x (|t| + m_{q^*}^2)/|t|$.  
Then, the differential cross section 
	for inelastic $e^+p$ scattering including 
	the effect of $q^*$ productions 
	can be calculated by using Eq.~(\ref{eq:1}) 
	with the following term added to $d\sigma $:
\begin{eqnarray}
	\Delta (d\sigma ) = 
	f_a (\xi ) d\xi {d \widehat \sigma _a\over dt},
\label{eq:13}
\end{eqnarray}
where
\begin{eqnarray}
	{d \widehat \sigma _a\over dt} = {   4 \pi \alpha ^2 \lambda _{q^*}^2
\over 
                               m_{q^*}^2 |t|      } 
	\left[ 1-y+ {       y^2 m_{q^*}^2
 \over 
		     2(|t|+m_{q^*}^2)   }\right] .
\label{eq:14}
\end{eqnarray}
For having 10 events for $|t| \geq  20000 {\rm GeV}^2$, 
	the H1 and ZEUS results \cite{1,2} 
	indicate $m_{q^*} \leq  136 {\rm GeV}$ for $\lambda _{q^*}=1$ 
	(95\% C.L.) 
	while having 313 events for $|t| \geq  5000 {\rm GeV}^2$, 
	they suggest $m_{q^*} \geq  116 {\rm GeV}$ 
	for $\lambda _{q^*}=1$ (95\% C.L.).
In Fig.\ \ref{f1}, the ratio $r$ including the effect of $q^*$ production 
	for $\lambda _{q^*}=1$ and for $m_{q^*}=116 {\rm GeV}$ 
	is compared with the H1 result \cite{1}.  

The sixth possibility is the production of excited positrons.  
For simplicity, let us assume that excited positrons 
	be produced electromagnetically via the effective interaction 
	whose Lagrangian is given by \cite{19}
\begin{eqnarray}
	L_{e^*} = {-e \lambda _{e^*}\over 2m_{e^*}}
 	\overline {e^*} \sigma _{\mu \nu } F^{\mu \nu } e + {\rm H.c.},
\label{eq:15}
\end{eqnarray}
	where $m_{e^*}$ is the excited electron mass and 
	$\lambda _{e^*}$ is a constant parameter, 
	and that they decay predominantly into 
	an electron and a composite boson ($X$) 
	via the effective interaction whose Lagrangian is given by \cite{20}
\begin{eqnarray}
L_X = f_{e^*} \overline {e^*} Xe + {\rm H.c.},
\label{eq:16}
\end{eqnarray}
	where $f_{e^*}$ is the coupling constant.  
	The decay width of $e^*$ can be calculated to be
\begin{eqnarray}
	\Gamma _X = {f_{e^*}^2 m_{e^*}\over 16 \pi } 
		\left( 1- {m_X^2\over m_{e^*}^2}\right) ^2,
\label{eq:17}
\end{eqnarray}
where $m_X$ is the $X$ mass.
The differential cross section for the process 
	of $e^+p \rightarrow e^{*+}p \rightarrow e^+Xp$ 
	can be calculated in the equivalent-photon approximation \cite{21} as
\begin{eqnarray}
	{d \sigma \over dt} \cong 
	{      4 \pi \alpha ^2 \lambda _{e^*}^2
\over 
	m_{e^*}^2 (m_{e^*}^2 - m_X^2)   }
	\left[ 1- {m_{e^*}^2\over s} + {1\over 2} \left( {m_{e^*}^2\over s}\right) ^2\right] 
	\ln \left[ {s(s-m_{e^*}^2)^2\over m_p^2 m_{e^*}^4} \right] .
\label{eq:18}
\end{eqnarray}
Let us define the invariant mass squared of 
	$e^{*+}$ channel $\tilde {s}$ by $\tilde {s} = (|t|+m_X^2)/y$.  
Then, the double differential cross section, 
	which will be needed in later discussions, 
	can also be approximated as
\begin{eqnarray}
	{d \sigma \over dxdt} \cong {
 			m_{e^*}^3 \Gamma _{e^*}
\over 
	\pi x[(\tilde s - m_{e^*}^2)^2 + m_{e^*}^2 \Gamma _{e^*}^2]} 
	{d \sigma \over dt},
\label{eq:19}
\end{eqnarray}
	where the terms vanishing at $\tilde {s} = m_{e^*}^2$ 
	are neglected.  
For having 10 events for $|t| \geq  20000 {\rm GeV}^2$, 
	the H1 and ZEUS results \cite{1,2} 
	indicate $m_{e^*} \leq  373 {\rm GeV}$ 
	for $\lambda _{e^*} = 1$ and $m_X = 100 {\rm GeV}$ (95\% C.L.) 
	while for having 313 events for $|t| \geq  5000 {\rm GeV}^2$, 
	they suggest $m_{e^*} \geq  304 {\rm GeV}$ for $\lambda _{e^*} = 1$ 
	and $m_X = 100 {\rm GeV}$ (95\% C.L.).
In Fig.\ \ref{f1}, the ratio $r$ including the effect of $e^{*+}$ productions 
	for $\lambda _{e^*} = 1$, $m_{e^*} = 304 {\rm GeV}$, 
	and $m_X = 100 {\rm GeV}$ is compared with the H1 result \cite{1}.  

To sum up, we have shown that the significant excess 
	found by the H1 and ZEUS Collaborations can be explained 
	by either one of the following possible consequences 
	of the substructure of quarks and leptons:
1) production of leptoquarks whose masses are around $300{\rm GeV}$, 
2) exchange of an excited $Z$ boson whose masses are around $200 {\rm GeV}$, 
3) intrinsic form factors of quarks (or leptons) 
	whose mass scales are around $500 {\rm GeV}$,
4) intrinsic anomalous magnetic moments of quarks whose mass scales 
	are around $150 {\rm GeV}$,
5) production of excited quarks whose masses are around $130 {\rm GeV}$, and
6) that of excited positrons whose mass is around $350$ {\rm GeV}.
Among these possibilities, the possibilities 2) and 5) contradict 
	with the currently available experimental bound of 
	$m_{Z'} > 505 {\rm GeV}$ 
	and $m_{q^*} > 570 {\rm GeV}$ (95\% C.L.) 
	(unless $q^*$ decays predominantly into channels 
	other than $qW$, $q \gamma $, and $qg$) \cite{22} 
	but the others may not be excluded 
	by any existing experimental constraints.  
In order to distinguish between these possibilities, 
	it may be extremely useful to investigate 
	not only the $t$-distribution of 
	events but also the others such as the $x$- and $y$-distributions.  
In Fig.\ \ref{f2}, we illustrate our predictions for the $y$-distributions 
	of the relevant ratio $r$ with the minimum $x$ cut of $x \geq  0.36$ 
	for various possibilities.  
The Figure shows 
	that the different possibilities may be well distinguished 
	in the $y$-distribution if a better statistics becomes available 
	in the coming H1 and ZEUS experiments.

In conclusion, in our two analyses of the CDF and HERA anomalies,
	we have found that possible production of excited bosons
	(excited gluons, weak bosons, Higgs scalars, etc.)
	whose masses are around 2TeV may be responsible for the CDF anomaly
	while that of either leptoquarks whose masses are around 300GeV
	or excited positrons whose masses are around 350GeV
	may be responsible for the HERA anomaly, 
	and that these two anomalies can be explained by different 
	``sectors'' of the same model, the composite model of
	not only quarks and leptons, but also gauge bosons and Higgs
	scalars \cite{6,7,8}.

\section *{Acknowledgements}
The authors would like to thank Dr.~Y.~Iga and Dr.~K.~Tokushuku 
	for giving them the valuable information on the ZEUS experiment.
They also wish to thank Dr.~I.~Ito for informing them of some
	numerical errors in the original manuscript.

\newpage


\epsfxsize=18cm\epsffile{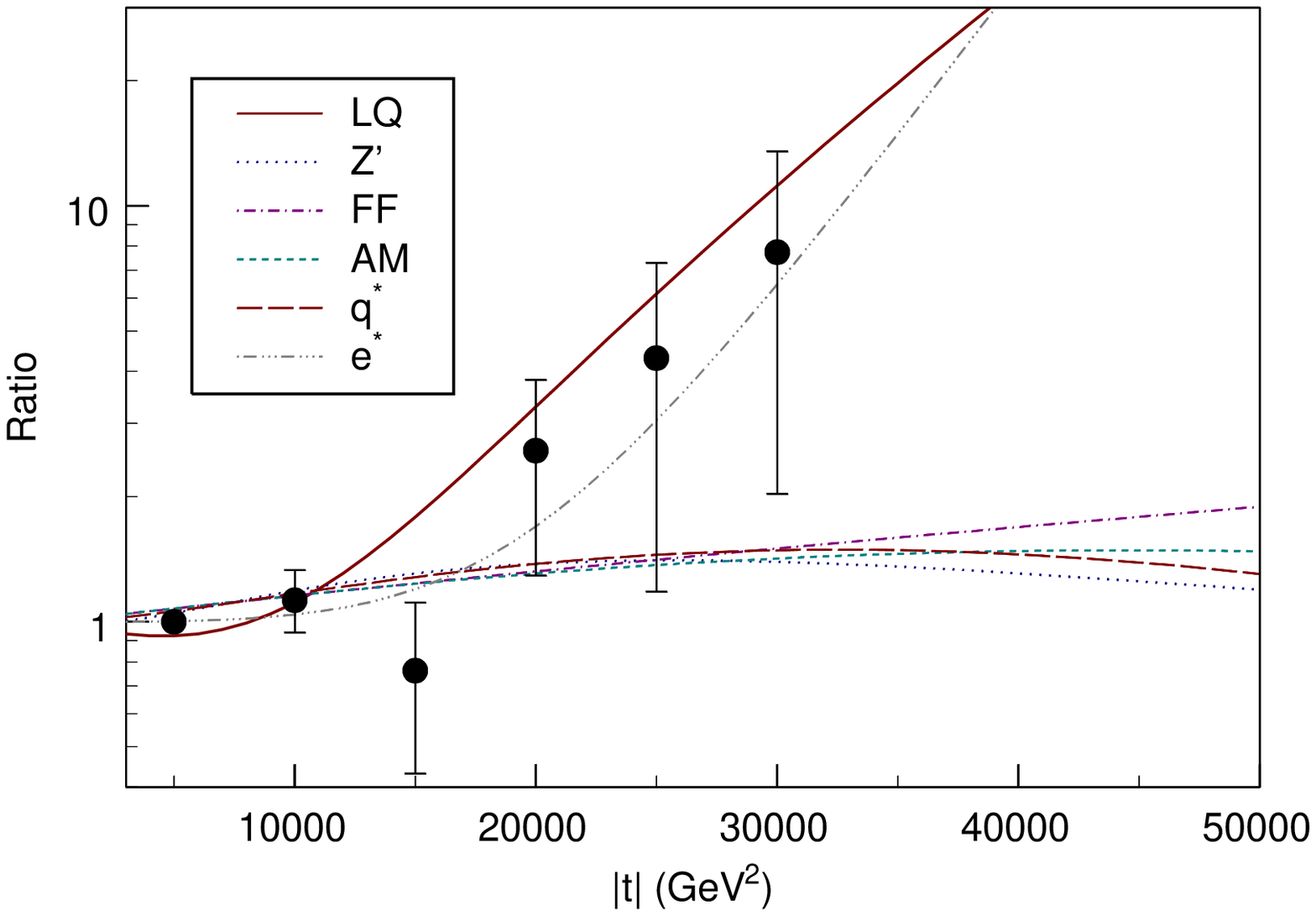}
\begin{figure}
\caption{ 
The predictions of the ratio of the differential cross section 
	$d\sigma /dt$ for inelastic $e^+p$ scattering 
	including one of the following effects 
	to that in the standard model:
the effects of 
1) leptoquarks (LQ), 
2) an excited $Z$ boson ($Z'$), 
3) intrinsic form factors of quarks(FF), 
4) intrinsic anomalous magnetic moments of quarks (AM), 
5) excited quarks ($q^*$), and 
6) an excited positron ($e^*$).  
The mass scale parameters are fixed at the 95\%C.L.\ lower bound 
	extracted by using the number of events 
	observed in the whole region of $|t|\geq 5000$GeV$^2$.
The points with an error bar 
	are the corresponding H1 experimental results 
	reported in Ref.\ [1].
}
\label{f1}
\end{figure}

\epsfxsize=18cm\epsffile{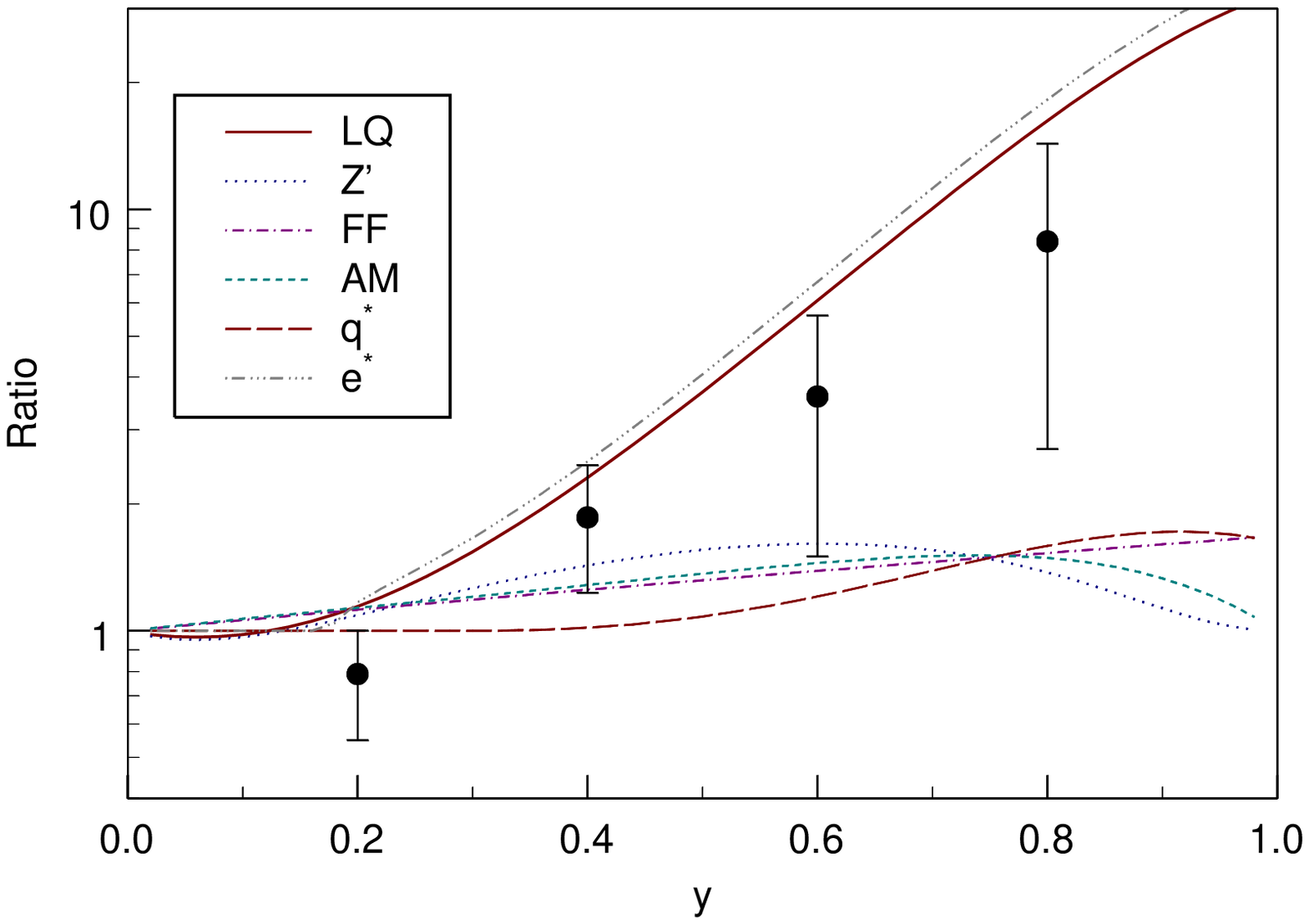}
\begin{figure}
\caption{ 
The predictions of the ratio of the differential cross section 
	$d\sigma /dy$ (with the minimum $x$ cut of $x \geq  0.36$) 
	for inelastic $e^+p$ scattering  
	including one of the following effects 
	to that in the standard model:
the effects of 
1) leptoquarks (LQ), 
2) an excited $Z$ boson ($Z'$), 
3) intrinsic form factors of quarks(FF), 
4) intrinsic anomalous magnetic moments of quarks (AM), 
5) excited quarks ($q^*$), and 
6) an excited positron ($e^*$).  
The mass scale parameters are fixed at the 95\%C.L.\ lower bound 
	extracted by using the number of events 
	observed in the whole region of $|t|\geq 5000$GeV$^2$. 
The points with an error bar are the corresponding H1 experimental results
	reported in Ref.\ [1].
}
\label{f2}
\end{figure}

\end{document}